\documentclass{aa}
\usepackage{graphicx}
\usepackage{amsmath}
\RequirePackage{natbib}
\begin{document}
\title{A photospheric bright point model}
\titlerunning{A photospheric bright point model}
\author{S.~Shelyag, M.~Mathioudakis, F.P.~Keenan, D.B.~Jess}
\authorrunning{S.~Shelyag et al.}
\institute{Astrophysics Research Centre, School of Mathematics and Physics, Queen's University, Belfast,
BT7 1NN, Northern Ireland, UK}

\date{01.01.01/01.01.01}

\abstract {}
{A magneto-hydrostatic model is constructed with spectropolarimetric properties close 
to those of solar photospheric magnetic bright points.}
{Results of solar radiative magneto-convection simulations are used to produce 
the spatial structure of the vertical component of the magnetic field. The horizontal 
component of magnetic field is reconstructed using the self-similarity condition, while 
the magneto-hydrostatic equilibrium condition is applied to the standard photospheric 
model with  the magnetic field embedded. Partial ionisation processes are found 
to be necessary for reconstructing the correct temperature structure of the model.}
{The structures obtained are in good agreement with observational data.
By combining the realistic structure of the magnetic field with the 
temperature structure of the quiet solar photosphere, 
the continuum formation level above the equipartition layer can be found. Preliminary 
results are shown of wave propagation through this magnetic structure. The observational  
consequences of the oscillations are examined in continuum intensity and in 
the Fe I 6302\AA\ magnetically sensitive line.}{}


\maketitle

\section{Introduction}
The study of magnetic elements at very small scales is one of the most important topics in solar physics. 
Magnetic Bright Points (MBPs) are ubiquitous in the solar photosphere. They have small diameters, 
typically less than 300 km, and are found in the intergranular lanes.  MBPs correspond to areas of 
kilogauss fields, are best observed in G-band disk centre images and are numerous in active regions 
or near sunspots. They are formed by a complex process involving the interaction of the magnetic 
field with the convectively unstable hot plasma. The physical processes associated with their formation 
have been outlined in \citet{shelyagbp1,shelyagbp2,carlsson}, using forward modelling of radiative 
magneto-convection in the solar photosphere and upper convection zone.

Despite the overall success of photospheric and sub-photospheric radiative magneto-convective models 
to reproduce many of the observational properties of solar radiation, we still do not fully understand 
the physical processes involved in the strongly magnetised photospheric plasma. In particular, it is 
difficult to use the results of these simulations for studies of acoustic wave propagation through 
the solar atmosphere and interior. Strong convective motions of the photospheric plasma can hide 
the signatures of acoustic waves, making them a difficult subject in both numerical and observational 
investigations.

The development of new methods for inferring the properties of solar plasma using sound waves 
have been followed by the successful modelling of the  magneto-acoustic properties in the solar 
atmosphere and interior \citep{hasan2, hanasoge, hasan1, fedun, khomenko2, parchevskii, shelyag4, 
steiner1,vigeesh1}. However, due to the non-locality of radiative processes in the solar atmosphere, 
a direct comparison of the plasma parameters at a certain {\it geometrical} depth in the computational 
box with the solar radiation parameters at a given {\it optical} depth may not be entirely correct. The 
non-locality of radiative transport must be taken into account.  \citet{khomenkol2009} have recently 
suggested that the changes in the height of continuum formation with respect to the equipartition 
layer (the layer where the Alfv{\'e}n speed is equal to the sound speed, $v_A=c_s$), may help to 
explain the appearance of the high-frequency acoustic haloes around  sunspots. 

Numerical simulations of solar wave phenomena require a static magnetic configuration model 
which incorporates as many physical properties of the real Sun as possible. In this paper we provide a recipe to create 
such a model, based on the results of numerical modelling of magneto-convection in the photosphere. 
We demonstrate that the spectropolarimetric properties of the magnetic configuration we created 
successfully reproduces those of MBPs. In Section 2 we describe the technique used to 
reconstruct the magnetic and thermal parameters of the MBP model. The results of the 
spectropolarimetric simulations using the model are presented in Section 3. In Section 4 we show our 
preliminary results on the wave mode conversion in the MBP and discuss the possible observational 
signatures.  Our concluding remarks are presented in Section 5.

\section{Static bright point model}

\begin{figure}
\includegraphics[width=1.0\linewidth]{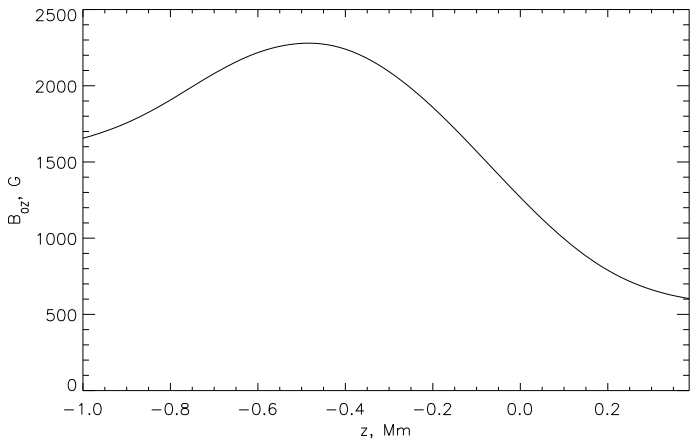}
\caption{Vertical magnetic field component $B_{0z}$ as a function of depth for the photospheric MBPs. A depth 
level of $z=0$ corresponds to $\log \tau_{4300}=0$ in the quiet solar model.}
\label{fig1}
\end{figure}

We use a snapshot from the "plage" magneto-convection simulation of the solar photosphere undertaken with
the MURaM code \citep{voegler1} to produce the average MBP model. Since the average 
magnetic field flux in this snapshot is relatively high ($200~\mathrm{G}$), a large number of 
intergranular magnetic field concentrations are generated. The magnetic field concentrations appear 
bright in the continuum ($\lambda=4300\mathrm{\AA}$), and in the G band. This snapshot has been 
used to demonstrate the magnetic nature of the G-band bright points \citep{shelyagbp1,shelyagbp2}. 
To reveal the basic structure of the vertical component of the magnetic field $B_{0z}$ in the MBPs, we 
average the depth dependences of the vertical magnetic field strength over the bright points, which 
are selected by their enhanced G-band intensity and magnetic field strength. Fig.~\ref{fig1} shows 
$B_{0z}$ as a function of depth. Note that $z=0$ on the depth scale corresponds to the average optical 
depth $\log \tau_{4300}=0$ for the quiet Sun, and thus does not include Wilson depression \citep{wilson}. 
The Wilson depression is about $100~\mathrm{km}$, as will be shown later.

\begin{figure*}
\includegraphics{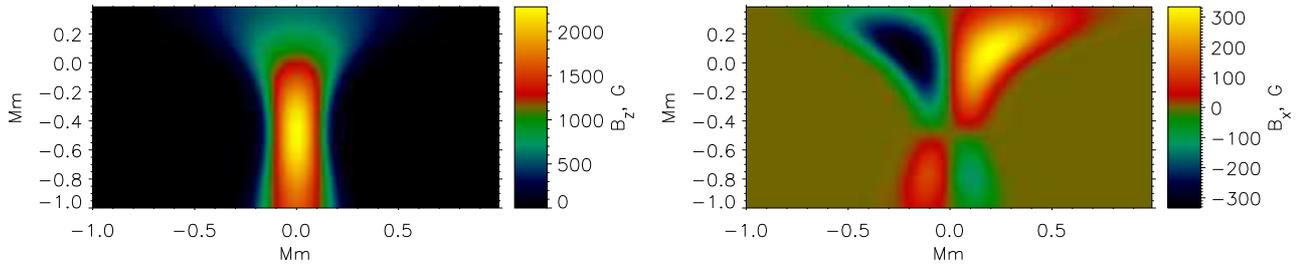}
\caption{Vertical (left) and horizontal (right) components of the reconstructed magnetic field for the MBP model.}
\label{fig2}
\end{figure*}

We use the self-similarity assumption \citep{schluter,schremp,gordovskyy,shelyag4} to reconstruct the 
horizontal component of the magnetic field. Below we briefly discuss its governing equations. 

For the prescribed vertical magnetic field component 
$B_{0z}\left(z\right)$, defined along 
the axis of the symmetric magnetic field configuration, the (divergence-free) two-dimensional magnetic field
can be defined as:
\begin{equation}
B_z\left(x,z\right)=B_{0z} f\left( x B_{0z} \right)
\label{eq1}
\end{equation}
and
\begin{equation}
B_x\left(x,z\right)=-\frac{\partial B_{0z}}{\partial z} x f\left( x B_{0z} \right),
\label{eq2}
\end{equation}
where $f$ is an arbitrary function which describes how the vertical component of the field expands with height. 

For the simulations, we choose the following grid parameters of the numerical domain in which the magnetic 
field is embedded: the vertical and horizontal extents of the domain are $1.4~\mathrm{Mm}$ and 
$2~\mathrm{Mm}$, respectively, which are resolved in $100\times200$ grid cells.

In Fig.~\ref{fig2} we show the magnetic field configuration calculated for this domain using the above definition. 
The function $f$ takes zero values on the side boundaries of the domain and is chosen in such a way that the 
radius of the magnetic field region is approximately $200~\mathrm{km}$ at the height corresponding to the 
quiet Sun continuum formation level.

The magnetic field changes the thermodynamic parameters of the plasma where it is embedded. These 
changes can be quite significant for the plasma dynamics and wave propagation (see e.g. \citet{shelyag4}). 
To construct the magneto-hydrostatic configuration, we substitute the values of the magnetic field 
components obtained using Eqs.~(\ref{eq1}) and (\ref{eq2}) into the equation of magneto-hydrostatic equilibrium 
\begin{equation}
\left({\bf B} \cdot \nabla\right){\bf B} + \nabla \frac{B^2}{2}+\nabla p = \rho {\bf g}.
\label{eq3}
\end{equation}
Here the magnetic field strength is normalised by the factor $\sqrt {4\pi}$. If the magnetic field is prescribed, 
Eq.~(\ref{eq3}) splits into a system of first-order partial differential equations describing the horizontal and vertical pressure and density 
variations from their equilibrium values. These equations are then integrated numerically. This procedure 
of integration gives satisfactory results in terms of numerical precision. 
%we do not use a more complicated method described by \citet{schremp}.

We use the standard model of the solar photosphere \citep{spruit} as the unperturbed background model.
The unperturbed pressure is recalculated from the standard density profile using the hydrostatic
equilibrium condition $\nabla p_u=\rho_u {\bf g}$. Then the perturbations to the pressure $p$ and density 
$\rho$, obtained from the solution of Eq.~(\ref{eq3}), are added to the unperturbed density and pressure
dependencies. 

\begin{figure*}
\includegraphics{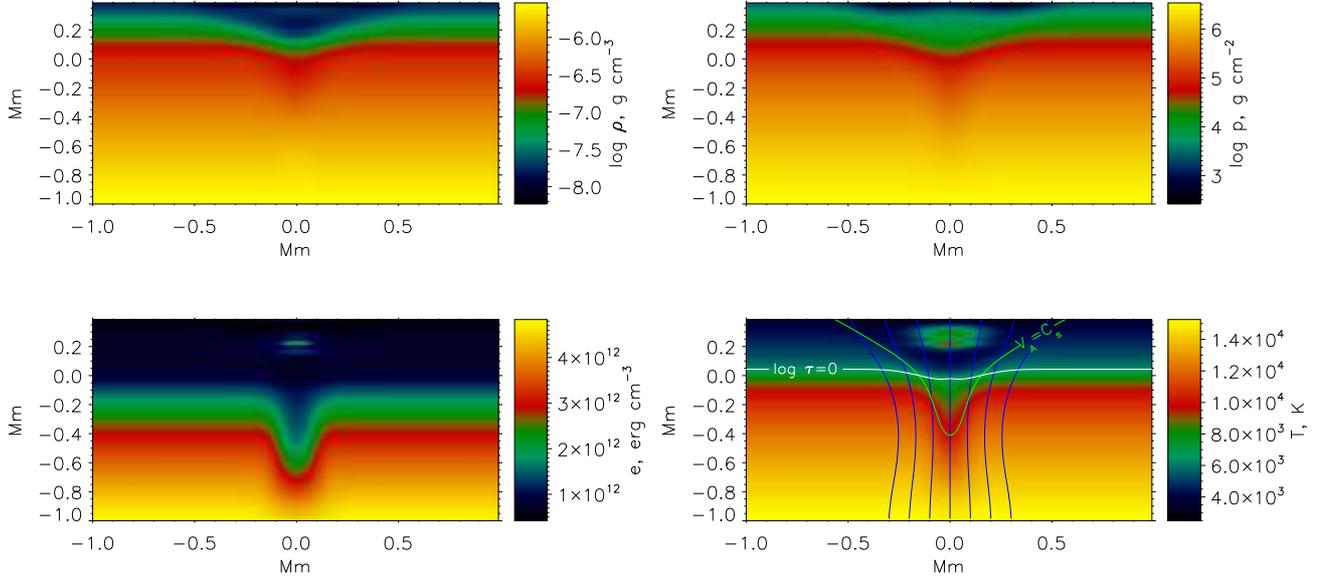}
\caption{The theoretical density (top left), pressure (top right), internal energy per unit volume (bottom left) and temperature 
(bottom right) structures of the MBP. In the bottom right panel, the white line shows the 4300\AA~continuum 
formation level $\log \tau_{4300}=0$, the green line represents the equipartition layer $v_A=c_s$, and the blue 
lines are the magnetic field lines.}
\label{fig3}
\end{figure*}

Partial ionisation effects are taken into account when the plasma internal energy and temperature are
calculated. The system of Saha equations is solved for the eleven most abundant elements in the solar photosphere 
to produce the tabulated functions of internal energy $e=e(\rho,p)$ and temperature $T=T(\rho,p)$. Values of 
internal energy and temperature are then obtained by interpolation. The density, pressure, internal energy and 
temperature structures are shown in Fig.~\ref{fig3}, where the lower right panel shows a significant temperature 
increase above the MBP. This temperature increase is caused by the increased magnetic tension in the upper 
layers. A similar, but somewhat weaker, temperature increase is observed in the dynamic radiative 
magneto-convection simulations.

For a direct comparison of the radiative properties of the model with observations, we need to know the position 
of the continuum formation at some wavelength relative to the equipartition layer ($v_A=c_s$). The thermal and 
magnetic structure of the average MBP is shown in the lower right panel of Fig.~\ref{fig3}, along with the equipartition layer 
and continuum formation level $\log \tau_{4300} = 0$. The latter curve demonstrates the presence 
of Wilson depression of the order of $100~\mathrm{km}$ in the magnetised region, similar to that observed for 
sunspots. The Wilson depression in the MBP model is significantly less than that of sunspots \citep{solanki2, watson}. The equipartition layer, 
as is evident from the figure, is located deeper than the continuum formation layer in the region of the strongest 
magnetic field. This means that in the MBP centre, the 4300\AA~continuum is originating from strongly 
magnetised plasma with $v_A > c_s$ and $\beta < 1$, and a variety of MHD effects, including the magneto-acoustic 
mode conversion, may be observable at this wavelength.

\section{Radiative diagnostics}
 
\begin{figure}
\includegraphics[width=1.0\linewidth]{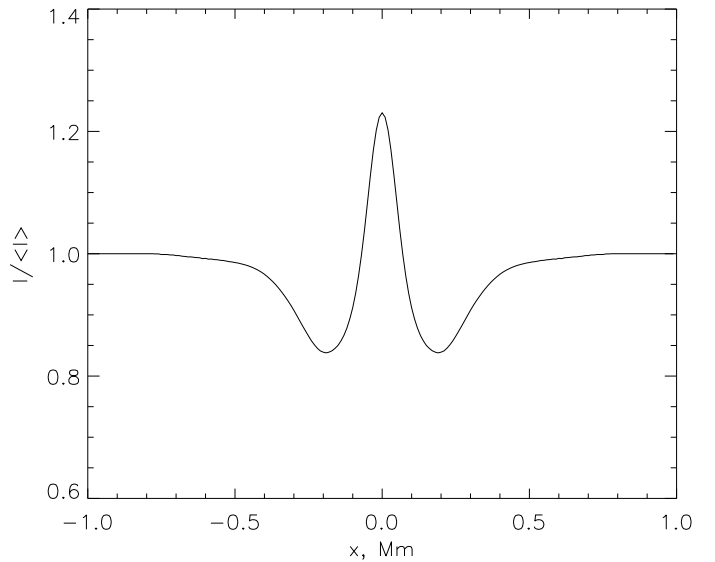}
\caption{G-band intensities for the average MBP model. The darkenings, corresponding to the integranular lanes, 
are found at $x=-0.3~\mathrm{Mm}$ and $x=0.3~\mathrm{Mm}$.}
\label{fig5}
\end{figure}

The line profile simulation code STOPRO \citep{solanki,frutiger,berdyugina,shelyagbp2,shelyag2} is used 
to perform the line profile calculations. This code is designed to compute wavelength-dependent intensities 
and normalised full Stokes vectors for the atomic and molecular line profiles in the LTE approximation.
We use STOPRO to calculate the continuum intensities at 4300\AA~and 5000\AA, as well as G-band intensities 
for each of the 200 vertical rays of the MBP model described in Section 2. To calculate these we employ a procedure 
similar to that described in detail in \citet{shelyagbp2}. The G-band intensities are derived by convolving the 
simulated spectrum with a 10\AA~filter, centered at 4305\AA, and integrating this over the range 4295\AA~- 
4315\AA. The intensities are shown in Fig.~\ref{fig5}, and have been normalised by the values 
taken in the non-magnetic part of the domain corresponding to granulation. As is evident from the figure, the 
MBP model appears to be bright in its middle $x=0$. Darkenings, corresponding to the integranular lanes 
surrounding the MBPs, are found at $x=-0.3~\mathrm{Mm}$ and $x=0.3~\mathrm{Mm}$.

\begin{figure}
\includegraphics[width=1.0\linewidth]{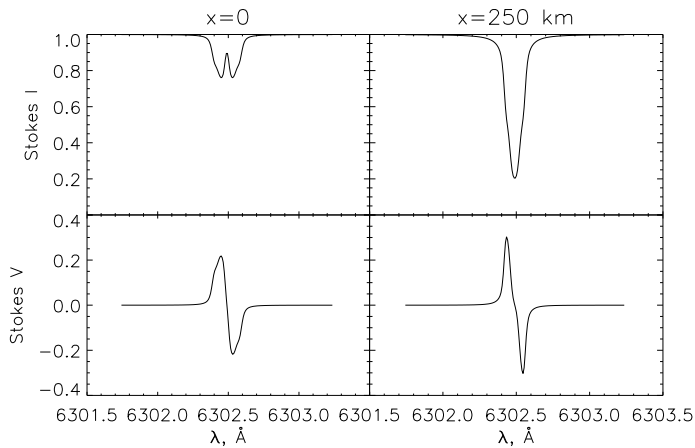}
\caption{Stokes-I (top) and -V (bottom) profiles calculated for magnetic flux concentration at $x=0$, (left), and 
$x=250~\mathrm{km}$, (right).}
\label{fig6}
\end{figure}

The 6302.49\AA~ Fe I transition is often used as a diagnostic for the photospheric magnetic field, and the Stokes-I and -V profiles
for this line are shown in Fig.~\ref{fig6} for two positions in the model. The I profile is split by strong magnetic field 
in the centre of the MBP, but the depth of the profile is reduced due to a smaller temperature gradient. In the weaker 
magnetic field (Fig.~\ref{fig6}, right column), the absorption is stronger, and the Stokes-V profile has a larger amplitude. However, 
the wavelength separation between the left and right lobes of Stokes-V remain approximately the same in both cases. It should 
be pointed out that the observational profiles are also influenced by the Doppler shift, caused by the bulk motions of the 
convecting plasma. Our model does not account for convective velocities and thus cannot reproduce the exact central 
wavelength and C-shape of the observed absorption line profiles.

\section{Wave propagation in the MBP}

We use the code SAC to perform simulations of wave propagation and mode conversion in the MBP 
\citep{shelyag4,fedun}. The code solves the full ideal MHD equations in a two- or three-dimensional Cartesian grid, 
and has previously been used for a variety of problems in the physics of solar oscillations. 
The boundary conditions implemented in the code are "transparent" conditions of the simplest type. These conditions lead to an artificial reflection from the boundary layers, however, for well-resolved perturbations the 
amplitudes of reflected waves are quite small. The full description of the methods used, boundary conditions and tests are presented in detail by \citet{shelyagcode}.
For the simulations, the code has 
been modified to incorporate the equation of state and account for partial ionisation processes. At each time step, the 
local internal energy- and density-dependent adiabatic index $\Gamma_1$ is calculated using the pre-calculated tabulated 
function $\Gamma_1=\Gamma_1\left( \rho, e\right)$. The adiabatic index is then used in the equation of state to relate the 
kinetic pressure to the internal energy values, thus closing the system of MHD equations.

Oscillations are generated using a single temporally continuous acoustic source located in the non-magnetic part of the 
numerical domain. This source is coherent, acts in the horizontal direction $v_x$, extents vertically over the whole height of 
the domain, and its period is $T=30~\mathrm{s}$.
The amplitude of the source is $100~\mathrm{m~s^{-1}}$, is constant over the domain depth and is chosen to be 
sufficiently small to keep the oscillations in 
the linear regime. The diffusive damping introduced into the code by increasing the hyperdiffusion coefficients to 0.2, 
combined with the low source amplitude, prevent the convectively unstable domain from going into the convective regime 
for a sufficiently long time.

\begin{figure}
\includegraphics[width=1.0\linewidth]{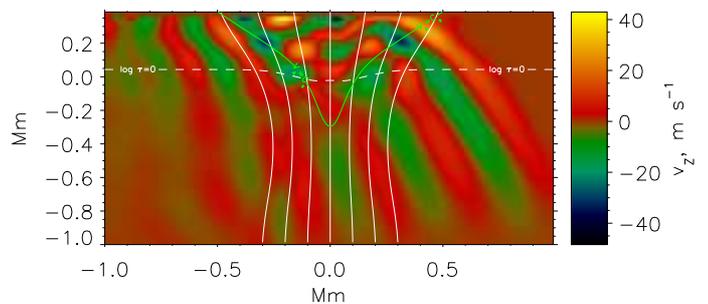}
\caption{Vertical velocity component in the domain. Lines as in lower right panel of Fig.~\ref{fig3}.
A complicated velocity pattern, produced by the interaction of the oncoming wave with the magnetic field concentration,
is visible above the equipartition layer $v_A=c_s$. }
\label{fig7}
\end{figure}

A snapshot of the vertical velocity component, taken at $t=160~\mathrm{s}$ from the start of the simulation, is shown 
in Fig.~ \ref{fig7}. An oscillatory pattern, bounded by the equipartition level $v_A=c_s$ (marked by the green line in 
the figure), is easily distinguishable. This pattern is produced by an interference of the slow and fast 
magneto-acoustic waves, propagating differently in the region. Most of the velocity pattern is above the 
continuum formation level $\log \tau = 0$ (white dashed line), and should therefore affect the absorption line profiles. 

\begin{figure}
\includegraphics[width=1.0\linewidth]{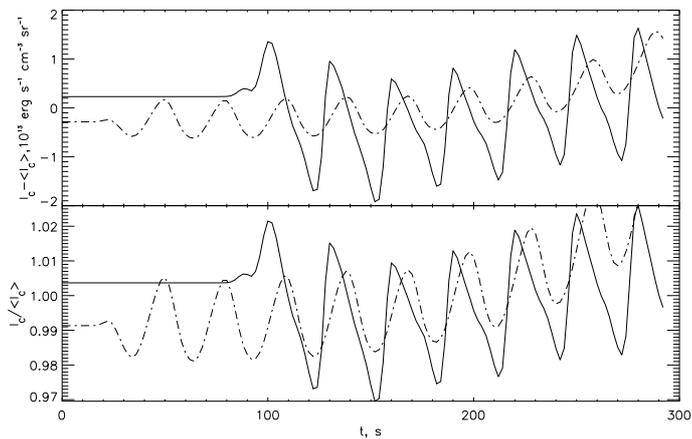}
\caption{Absolute (upper plot) and relative (lower plot) continuum intensity oscillations in the non-magnetic media 
($x=0.66~\mathrm{Mm}$, dashed line) and in the bright point centre ($x=0$, solid line).}
\label{fig8}
\end{figure}

Fig.~\ref{fig8} shows the 5000\AA~continuum intensity variations in non-magnetic media outside the MBP
(dashed lines) and in the centre of the MBP (solid lines). It is evident from the figure that both the absolute 
intensity variation (upper plot) and the intensity variation normalised to the average continuum 
intensity (lower plot), are larger in the MBP than in non-magnetic media. This is caused by the larger 
continuum intensity and stronger variation of temperature in the deeper (due to depression) layers of the solar photosphere 
in the MBP.  Although the source amplitude is sufficiently low to keep the linear character of the
intensity oscillations in the non-magnetic part of the domain (dashed lines), a non-linear character of the intensity oscillations in
the MBP is also observed (solid lines). The steepening of the wave front observed in the figure 
is caused by conversion of the linear wave into the non-linear regime in the partially evacuated plasma of the MBP.

\begin{figure}
\includegraphics[width=1.0\linewidth]{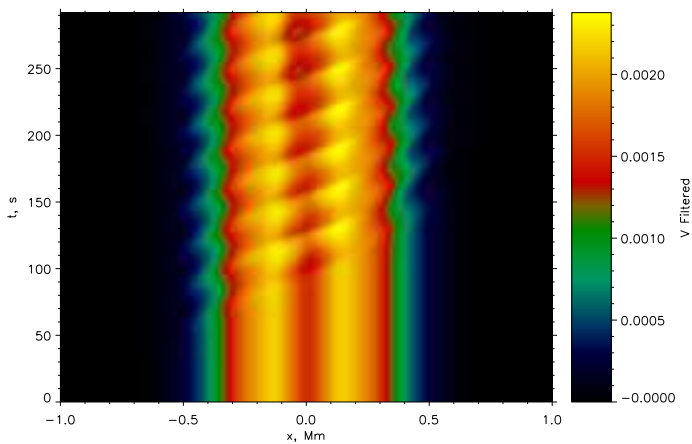}
\caption{Filtered Stokes-V intensities, calculated for the simulated time series.}
\label{fig9}
\end{figure}

An oscillatory behaviour can also be detected in the magnetic field. To reveal the observational consequences of the 
oscillations in the magnetic field, we apply the standard observational technique of measuring the Stokes-V profile. 
We choose a narrow bandpass filter with a width of $0.02\mathrm{\AA}$, centered $-0.0525\mathrm{\AA}$ from the 
Fe I line centre. We then apply this filter to the Stokes-V profiles data, calculated for the whole $x-t$ domain. The 
resultant Stokes-V intensities are shown in Fig.~\ref{fig9}. The vertical component of the magnetic field in the region of the 
line formation is proportional to the intensity of the Stokes-V filter. However, it is worth noting that the Stokes-V profile for the 
6302\AA~line saturates in regions of strong magnetic field. As Fig.~\ref{fig9} demonstrates, the intensity  
at $x=0$ is equal to the intensity at about $\pm 0.3~\mathrm{Mm}$. Also, the magnetic field strength values, restored from 
the Stokes-V intensities at these positions, are virtually indistinguishable, thus making the 6302.49\AA\ Fe I line unsuitable 
for measurements of strong magnetic fields. The variation in the filtered Stokes V amplitudes in the centre of MBP is 
about 25-30\% in our simulations. This value is quite large and should therefore be detectable by observations.

\begin{figure}
\includegraphics[width=1.0\linewidth]{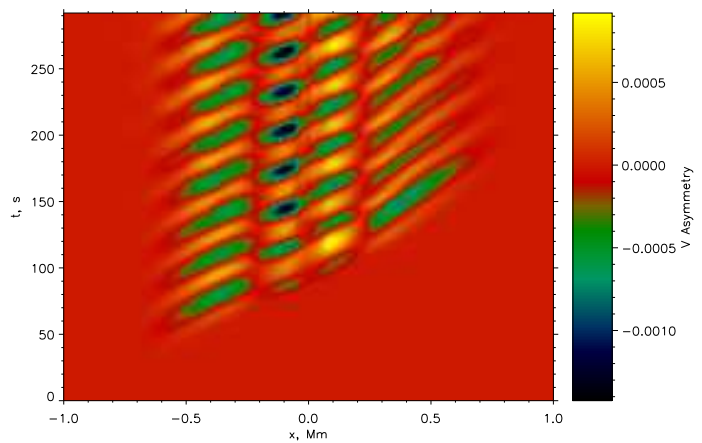}
\caption{Stokes-V asymmetries, calculated for the simulated time series.}
\label{fig10}
\end{figure}

The magnetic field configuration in the simulations is localised and non-uniform, and the line-of-sight therefore crosses 
regions of both zero and non-zero magnetic field. The initial static model does not produce an asymmetry in the Stokes-V profiles, 
since it has no velocities which are necessary to produce an asymmetry (see e.g. \citet{sankara,bellot1}). In the presence of a flow, caused by 
the perturbation introduced into the numerical box, this magnetic field configuration can lead to asymmetric Stokes-V 
profiles \citep{solanki2}. The simplest measure of the area asymmetry is the integral of the Stokes-V 
profile over the wavelength $A=\int V d\lambda$. The V profile area asymmetry map is shown in Fig.~\ref{fig10}.
The normalised amplitude of Stokes-V area asymmetry is of the order of 2\% in the center of the MBP. This value is close to the observed
values of Stokes-V area asymmetries, which are about 3\% \citep{bellot1}. Thus, the variation of the Stokes-V area asymmetry
produced by the perturbation is large enough to be detected by modern instruments.

\section{Concluding remarks}
We have presented the recipe for creating a static numerical model of a photospheric MBP.  
This model allows the study of oscillatory properties of such small magnetic configurations 
in the solar photosphere directly by measuring their radiation intensity and polarimetric 
properties. The magnetic field is extracted from dynamic simulations of solar radiative 
magneto-convection and reconstructed using the assumption of self-similarity. Thus, 
our static model inherits many observational properties of the full dynamic simulation.

The model reproduces the G-band brightening in the MBP centre. Stokes-I and -V 
profiles for the $6302\mathrm{\AA}$ FeI line are comparable to solar observations in 
both the magnetised region and in the region 
corresponding to the non-magnetic solar granulation. The continuum formation level in the 
centre of the MBP is located in the region of strongly magnetised plasma, where the 
Alfv{\'e}n speed is greater than that of sound.
 
We have used the model to examine the  observational consequences 
of sound wave propagation though the magnetic field concentration of the MBP. This preliminary 
investigation shows that the variation of continuum intensity is more pronounced in the MBP 
compared to the average granule. Using the radiative diagnostics with the Stokes-V profile of the  
$6302\mathrm{\AA}$ Fe I absorption line, we demonstrate the detectability of the magnetic field 
variation in the bright point and show the  appearance of the Stokes-V profile asymmetry 
caused by the oscillations.

The radiative heating term is not included into the system of MHD equations used to perform the
wave dynamics study. The absence of the heating term may result in reduced vertical radiative flux 
in the magnetic flux tube. As a result, we have somewhat smaller continuum and G-band intensity in the 
MBPs, compared to the full simulations and the real Sun. However, for wave dynamics the effect of 
the radiative term is expected to be small. As it has recently been shown by \citet{yelles}, the magnetic 
flux tubes in the solar photosphere are "reasonably well reproduced" by a thin flux tube approximation 
with no radiative term included.
 
Future investigations will examine the response to sources at different locations and with a range 
of frequencies. The behaviour and observational signatures of Alfv{\'e}n waves in small magnetic 
elements is particularly important when used as a diagnostic for solar plasma parameters, and 
for understanding the energy transport to the corona. Thus we plan to extend the model to 
three dimensions.

\section*{Acknowledgements}
This work has been supported by the UK Science and Technology Facilities Council (STFC). 
FPK is grateful to AWE Aldermaston for the award of a William Penney Fellowship.

\end{document}